# Teaching materials aligned or unaligned with the principles of the Cognitive Theory of Multimedia Learning: the choices made by Physics teachers and students

Aline N. Braga[1], Antonio A.M. Neto[1,2], Alessandra N. Braga[3], Silvio C.F. Pereira Filho[4], Nelson P.C. de Souza[5], Danilo T. Alves [1,6]

[1] Universidade Federal do Pará, Instituto de Educação Matemática e Científica, Belém, PA, Brasil.
[2] Universidade Federal do Sul e Sudeste do Pará, Instituto de Ciências Exatas, Marabá, PA, Brasil.
[3] Universidade Federal do Pará, Faculdade de Física, Ananindeua, PA, Brasil
[4] Universidade Federal do Pará, Campus Universitário do Marajó-Breves, Faculdade de Ciências Naturais, Breves, PA, Brasil.
[5] Universidade Federal do Pará, Escola de Aplicação, Belém, PA, Brasil.
[6] Universidade Federal do Pará, Faculdade de Física, Belém, PA, Brasil.

ABSTRACT:

In a recent study [Rev. Bras. Ens. Fís. vol. 45, 2023], the absence of the Cognitive Theory of Multimedia Learning (CTML) in the curricula of Physics teacher education programs at Brazilian public universities was highlighted. Considering this gap, the present study investigates whether, even without any formal prior knowledge of CTML principles (Coherence, Signaling, Spatial Contiguity, Segmentation, Multimedia, and Personalization), Physics teacher trainees and educators tend to choose, among two formats of multimedia materials—one aligned with a given CTML principle and the other not—the materials aligned with these principles. The findings of this case study revealed that, although most participants generally selected materials aligned with the mentioned principles, a significant portion did not. These results underscore the importance of Brazilian universities considering the inclusion of CTML in Physics teacher education curricula.

# 1 INTRODUCTION

The Cognitive Theory of Multimedia Learning (CTML), developed by American psychologist Richard Mayer, was created with the aim of understanding how meaningful learning occurs from materials involving words and images (Mayer, 2024). This theory is based on three fundamental assumptions about human learning, which are classic concepts of cognitive psychology: the existence of two main channels for processing information (visual/pictorial and auditory/verbal); the limited capacity of both to process information; and the assumption that such processing takes place actively, involving attention to important information, mental organization and the association of this information with previous knowledge, this being the fundamental point for CTML (Mayer, 2024).

Mayer began his studies on multimedia learning in 1989. Based on evidence, with more than 200 experiments carried out by him and his collaborators, CTML proposes 15 principles that guide the creation of teaching materials and educational practices with the aim of promoting greater learning (Mayer, 2021; Mayer, 2024). The development of this theory has been enhanced by research produced by laboratories in different parts of the world (Mayer, 2024).

The fundamental principle of CTML is the "Multimedia Principle", which postulates that "people learn better with words and images than with words alone" (Mayer, 2021, p. 117). The combination of words and images designed to facilitate learning is called a "multimedia instructional message" (Mayer, 2021). According to Mayer (2021), it is possible to use multimedia approaches even in low-tech contexts, such as in a lesson where the teacher writes or draws on the board while explaining. In this way, the teacher can combine words (written or spoken) and images (illustrations, graphs, photos) to improve students' understanding of the content being taught, rather than limiting themselves to the verbal transmission of information. This is because "when we present material only in the verbal mode, we are ignoring the potential contribution of our ability to process material in the visual mode as well" (Mayer, 2021, p. 7). According to Mayer (2021, p. 17), multimedia learning involves "an activity in which the learner seeks to construct a coherent mental representation from the material presented", in other words, a logical, consistent and organized representation of the material presented.

The principles of CTML have been extensively studied over the past 40 years, as evidenced by works such as Clark & Mayer (2011), Fiorella & Mayer (2015), Gemino, Parker & Kutzschan (2006), Johnson & Mayer (2012), Mayer & Johnson (2008), Mayer, Sobko & Mautone (2003), Mayer & Dapra (2012), and Moreno & Mayer (2000). In a bibliometric analysis study on trends and research issues in multimedia learning, Li, Antonenko, and Wang (2019) revealed that, over the last two decades, research in this field has focused on five main strands: theoretical foundations of multimedia learning, representations and principles, instructional design and individual differences, motivation and metacognition, and video and hypermedia.

In Brazil, recent work has applied CTML to optimize teaching materials, especially in analyzing the effectiveness of images and illustrations in textbooks. In addition, there is interest in applying CTML to the use of digital media and online platforms to improve teaching (Almeida, Chaves, Coutinho & Araújo Júnior, 2014; Alves & Ramos, 2016; Coutinho, Soares & De Moura Braga, 2010; Coutinho & Soares, 2010; Das Neves, Dos Anjos Carneiro-Leão & Ferreira, 2016; De Sousa, De Oliveira, De Oliveira Silva, & Das Neves, 2023; De Matos, Coutinho, Chaves, De Jesus Costa & Amaral, 2010; Martins, Da Conceição Galego & De Araújo, 2017; Menezes, 2021; Silva & Fonseca, 2020; Silva et al., 2020; Stafusa, Santos & Cardoso, 2020; Teodoro & Menezes, 2021).

Despite its relevance, CTML is not yet incorporated into the undergraduate physics curricula of Brazilian public universities, as revealed in a recent study (Neto et al., 2023) which identified this gap in physics teacher training. This suggests the need for a curriculum review to include CTML and improve the preparation of future physics educators. Furthermore, as discussed by Mayer in an interview (Torkar, 2022), teachers play a crucial role in the selection and effective use of multimedia instructional materials, which reinforces the importance of including CTML in academic training.

Colombo e Antonietti (2006) conducted the following study on the effectiveness of multimedia materials aligned or not with the principles of CTML. With a total of 112 participants, undergraduate students in Business, Psychology, Languages and Arts, the authors asked the participants to imagine that a university student would learn about the topic of "lightning" through the texts contained in the survey questionnaire. The participants were asked to give a score quantifying the efficiency of the learning produced by each material. In all, five pairs of materials were tested, with one pair aligned and the other not aligned with a given CTML principle. Five CTML principles were examined: Spatial Contiguity, Temporal Contiguity, Coherence, Modality and Redundancy. The authors concluded that participants always gave higher scores (agreement) to presentations aligned with the aforementioned CTML principles. They also pointed out that the vast majority of participants agreed with materials aligned with the principles of Spatial Contiguity, Temporal Contiguity and Coherence. With regard to the remaining principles,

Modality and Redundancy, the percentages of agreement with materials aligned with these principles were lower, but still higher than the percentages of disagreement. Specifically, in relation to the principles of Spatial Contiguity and Coherence, these authors obtained the following results: 88.4% and 69.0%, respectively, were the percentages of choices in agreement with these two principles (later on, we will compare these results with those obtained in this research).

Colombo e Antonietti (2006), However, they restricted their research to undergraduate students in Business, Psychology, Languages and Arts, and did not include undergraduate students in Physics, nor did they include teachers (the importance of including teachers will be discussed later). Nor did these authors investigate the principles of Signaling, Segmentation, Multimedia or Personalization. In this study, we expand on the research by Colombo and Antonietti (2006), considering physics teachers and undergraduate students as participants, and analyzing the principles of Signaling, Segmentation, Multimedia and Personalization, as well as analyzing two principles also considered by Colombo and Antonietti (2006): the Principle of Coherence and the Principle of Spatial Contiguity. We investigated whether, even without any prior formal knowledge of CTML, these participants choose between two formats of multimedia material, one aligned with a given principle of CTML and the other not, those aligned with these principles. We asked the participants in this study to select one of these two formats of material to use with their students in a hypothetical physics lesson. To conduct the research, six pairs of multimedia presentations were prepared (one pair for each principle). To avoid the results of this research being prejudiced by purely random selections, validation questionnaires were introduced to rule out any random selections, validating the non-random results. The results of the investigation (case study) revealed that although the majority of participants generally chose materials in formats in line with the aforementioned CTML principles, a significant proportion of participants did not. These results reinforce the importance of Brazilian universities considering the inclusion of CTML in physics teacher training curricula.

It is worth mentioning that, unlike Colombo and Antonietti (2006), this study also includes teachers as participants. The aim of including this group of participants is to see if the pedagogical knowledge accumulated through experience has an impact on the choice of material formats in line with the principles of CTML.

This paper is organized as follows. In Section 2, we discuss the human cognitive process, presenting CTML and its principles for multimedia learning. In Section 3, we present the research methodology, covering the procedures for its realization, the instruments used for data collection, the identification of the locus and the participants, as well as the application of the research. The aim is to make it clear how the data was collected. In Section 4, we present the analysis and discussion of the results, structured in three stages: first, we examine the participants' preferences in relation to the six pairs of multimedia presentations, identifying the number of participants who chose formats aligned with the principles of CTML; then, we report the number of participants who chose and justified their choices in a way that was aligned or not with the principles of CTML; finally, we deepen the analysis, interpreting the results of the participants' choices and justifications in the light of the Theory of Significant Learning. In Section 5, we present our final considerations.

## 2 COGNITIVE PROCESSING AND THE COGNITIVE THEORY OF MULTIMEDIA LEARNING

For CTML, the cognitive process during learning involves three types of memory, in line with Baddeley's (2009) memory model: sensory, working and long-term. Sensory memory is the initial stage of information processing in the cognitive structure and is responsible for connecting the stimuli captured by the senses to the cognitive structure. Although it has a great capacity for capturing stimuli, its retention is brief and limited for retrieving information. Working memory is responsible for storing visual/pictorial and auditory/verbal information captured by sensory memory. According to Baddeley, working memory is "a system of limited capacity that allows the temporary storage and manipulation of information required for complex tasks such as comprehension, learning and reasoning" (2000, p. 418).

Working memory is limited both in the length of time it takes to retain information, which is approximately 18 seconds (Peterson; Peterson, 1959; Solso, 1995), and in terms of capacity, being able to store up to 7±2 unrelated items simultaneously (Miller, 1994). However, Cowan found that this limit of 7±2 items only applies to simple tasks; for complex tasks, such as those involving multiple simultaneous activities, the limit is reduced to 4 items (Cowan, 2001; Mathy; Chekaf & Cowan, 2018). These limitations affect learning and should therefore also influence teaching approaches (Souza, 2010). Instructional strategies must therefore take these restrictions into account. Finally, long-term memory is the memory responsible for storing knowledge permanently, "... a storehouse of almost unlimited capacity. Like a new file that we save on a hard disk, the information in long-term memory is archived and encoded so that we can access it when we need it" (Feldman, 2015, p. 211).

CTML also describes five essential cognitive processes: word selection, image selection, word organization, image organization, and integration. These processes guide the structuring and connection of verbal and visual information, facilitating meaningful learning. Mayer (2024) also describes three types of processing that can occur during learning and which compete for working memory capacity: extraneous, which corresponds to cognitive processing that does not contribute to the learning objective, such as the inclusion of images or texts unrelated to the main content in teaching materials; essential, which comprises the cognitive processing needed to represent the material in working memory in the mind, and its occurrence is linked to the complexity of the material (for example, it can occur when introducing several complex concepts quickly in a single lesson); and generative, which consists of the cognitive process that seeks to make the information received meaningful, and the occurrence of this cognitive process is related to the level of motivation of the learner in striving to understand the material presented to them.

Mayer (2021) emphasizes the importance of counterbalancing the occurrence of these processes since, as mentioned, they compete for working memory capacity during exposure to multimedia materials, such as the use of textbooks. Thus, learning becomes efficient to the extent that extraneous cognitive processing can be reduced, essential cognitive processing can be managed and generative cognitive processing can be made viable. To this end, Mayer and his collaborators developed fifteen principles, which are shown in Figure 1.

**Figure 1. Principles of multimedia learning.**

1. Coherence principle: People learn better when extraneous material is excluded rather than included
2. Signaling principle: People learn better when cues are added that highlight the organization of the essential material
3. Redundancy principle: People do not learn better when printed text is added to graphics and narration; people learn better from graphics and narration than from graphics, narration, and printed text, when the lesson is fast-paced
4. Spatial contiguity principle: People learn better when corresponding words and pictures are presented near rather than far from each other on the page or screen
5. Temporal contiguity principle: People learn better when corresponding words and pictures are presented simultaneously rather than successively
6. Segmenting principle: People learn better when a multimedia lesson is presented in user-paced segments rather than as a continuous unit
7. Pretraining principle: People learn better from a multimedia lesson when they know the names and characteristics of the main concepts
8. Modality principle: People learn better from graphics and narration than from graphics and on-screen text
9. Multimedia principle: People learn better from words and pictures than from words alone
10. Personalization principle: People learn better from multimedia lessons when words are in conversational style rather than formal style
11. Voice principle: People learn better when the narration in multimedia lessons is spoken in a friendly human voice rather than a machine voice
12. Image principle: People do not necessarily learn better from a multimedia lesson when the speaker's image is added to the screen
13. Embodiment principle: People learn more deeply from multimedia presentations when an onscreen instructor displays high embodiment rather than low embodiment
14. Immersion principle: People do not necessarily learn better in 3D immersive virtual reality than with a corresponding 2D desktop presentation
15. Generative activity principle: People learn better when they are guided in carrying out generative learning activities during learning

**Source:** adapted from Mayer (2024, p. 20).

Despite the importance of CTML in guiding the effective selection, use and preparation of teaching materials - as mentioned above - it has not yet been incorporated into the undergraduate physics curricula of Brazilian public universities (Neto et al., 2023). This constitutes a significant gap in the training of physics teachers.

Among the factors that contribute to the inclusion of a theory in teacher training curricula in a given country is its dissemination through scientific publications and academic events. David Ausubel's Significant Learning Theory, for example, was introduced to Brazil "in the early 1970s by Prof. Jael Martins, when he began teaching postgraduate courses at the Pontifical Catholic University of São Paulo, based on the ideas of this American researcher" (Ronca, 1994, p. 91). Ausubel was also in Brazil in 1975, when he coordinated an Advanced Seminar that brought together 25 researchers from all over the country at PUC-SP. Since then, various research projects have begun to explore different aspects of Ausubel's theory, including its application to physics teaching (Ronca, 1994; De Jesus et al., 2022; Moreira, 2003; Moreira, 2021a; Moreira, 2021b).

Currently, CTML is in a similar condition to the initial period of Ausubel's theory in Brazil. Although Mayer's works on CTML have not yet been translated into Portuguese, the theory has already begun to be discussed at academic events and in scientific articles in the country, as presented in the previous section. This work also seeks to contribute to the dissemination of this theory in Brazil, encouraging its inclusion in physics teacher training curricula.

## 3. METHODOLOGY

### 3.1 General aspects

The aim of this study was to investigate whether physics undergraduates and teachers, even without prior formal knowledge of CTML, choose multimedia materials in line with CTML principles. To achieve this goal, six pairs of multimedia presentations were prepared, each pair intended for analysis corresponding to a CTML principle. Each pair consisted of one presentation aligned with a given principle and one that was not. The study's analysis focused on the following principles applicable to printed materials: Coherence Principle, Signaling Principle, Spatial Contiguity Principle, Segmentation Principle, Multimedia Principle and Personalization Principle. In addition to the presentations, validation questionnaires were developed for each pair, as well as a final questionnaire.

The research was carried out at the Federal University of Pará (UFPA), on the Ananindeua and Belém campuses. A group of 13 people took part in the research, five of whom were physics teachers and eight physics undergraduate students. From now on, the two subgroups of research participants will be referred to as follows: teachers trained in Physics (referred to as the "teachers' subgroup") and Physics undergraduate students (referred to as the "undergraduates' subgroup"). The data collection with the students took place in the subject of Fundamental Physics IV at the Ananindeua campus, while the data collection with the teachers was conducted at a meeting held at the Institute of Mathematical and Scientific Education at UFPA, Belém campus.

The study used a case study methodological approach, investigating both qualitative and quantitative aspects. A case study refers to a specific case, which must be relevant in a given circumstance in order to "support a generalization for analogous situations, authorizing inferences" (Severino, 2007, p. 121). The research used structured questionnaires to collect data, the answers to which were thoroughly analyzed to identify the participants' choices and justifications for multimedia presentation formats.

With regard to the analysis technique proposed for this research, we adopted some elements of the content analysis method (Bardin, 2011). We began with a pre-analysis of the material through an initial reading to categorize the responses according to the CTML principles investigated (Coherence, Signaling, Spatial Contiguity, Segmentation, Multimedia and Personalization). At this stage, we identified the participants' choices for each pair of materials (aligned or not with the principle) and their associated justifications. We then coded the data using the justifications for choosing the teaching material and the criticisms of the material not chosen. The coding allowed for a qualitative analysis of the justifications, highlighting the extent to which the participants' arguments were aligned with the CTML principles. After coding, the answers were validated according to pre-defined criteria to ensure that the participants' choices were in fact related to the CTML principles. This validation considered both the coherence of the justifications with the principles and the exclusion of answers that indicated random selections or lacked theoretical justification. Once the data had been validated, we proceeded to interpret the results. This interpretation focused on understanding the factors that influenced the choice of materials aligned or not with the principles of CTML. In addition, the analysis considered how the justifications given by the participants reflect their understanding, even without formal knowledge of the theory, using concepts from Ausubel's Meaningful Learning Theory to also interpret the results.

Below we describe the stages of the study.

*Stage 1*

The research began with the distribution of the Informed Consent Form, which the participants were asked to fill in. They were then told that their participation was voluntary. The following steps were then carried out.

*Stage 2*

Stage 2/1: Each participant was given a pair of multimedia presentations in printed form, one in a format aligned and the other not aligned with a given CTML principle. The participants were instructed to select one of the two formats of multimedia material to use in a hypothetical physics lesson, selecting the format they considered most beneficial for learning.

Stage 2/2: Participants were then asked to justify their choices and to point out flaws in the presentations not chosen (this was done using the validation questionnaire).

Stages 2/1 and 2/2 were repeated for the six CTML principles investigated here. This completes Stage 2.

*Stage 3*

The participants were asked to answer a final questionnaire, which collected information such as their name, age, stage of their degree course, whether they had teaching experience and, in the case of the teachers, whether they were also studying for a postgraduate degree. They were also asked if they had any prior knowledge of the principles of CTML.

*Stage 4*

The research sought to analyze and interpret the material collected in Stage 2 under the guidance of the following theoretical frameworks: Mayer's CTML (2021), which underpinned the entire research, from the initial data collection phase to part of the interpretation of the results; and Ausubel's Significant Learning Theory (2000), more specifically the concepts of subsumers and advanced organizers, which were also used to interpret the results.

## 3.2 Specific Aspects of Stage 2/1: Pairs of Multimedia Presentations

With regard to the construction of the pairs of presentations, these were based on adaptations of the book "Ciência Animada: uma introdução ao estudo da óptica" (Animated

Science: an introduction to the study of optics) (Costa, 2021), which is an educational product developed as part of the book's author's master's research at the Graduate Program in Teaching in Science and Mathematics Education at UFPA. This material is specifically aimed at teaching optics in the 9th year of elementary school (Costa, 2021).

In each multimedia presentation pair, one of the formats followed one of the CTML principles (Coherence, Signaling, Spatial Contiguity, Segmentation, Multimedia and Personalization), while the other did not. The information aligned and not aligned with the principles was organized into presentation formats named Format 1 and Format 2. These pairs had the same learning purpose, i.e. they addressed the same physics teaching content, differing only in the way the information was organized.

It's important to note that formats 1 and 2 were organized and applied randomly to avoid the research subjects being influenced by patterns. In the presentations on the principles of Coherence, Signaling and Spatial Contiguity, Format 1 was not aligned with these principles, while Format 2 was aligned. In the presentations on the principles of Segmentation, Multimedia and Personalization, Format 1 was aligned with these principles, while Format 2 was not. This organization allowed the research participants to evaluate the different multimedia presentation formats more effectively. Below, we will highlight the particularities of each presentation format, highlighting the elaboration according to the purpose of the analysis of the perception of each principle: Coherence, Signaling, Spatial Contiguity, Segmentation, Multimedia and Personalization.

***Coherence Principle: selection of presentation formats on the process of perceiving the color blue***

For the analysis of perception in relation to the Principle of Coherence, the pairs of multimedia presentations were designed with the aim of "*helping the student to understand the process of perceiving the color blue, through the phenomena of reflection and absorption of light*", as indicated to the participants. More specifically, to help students explore how the color of an object, in this case a table, is perceived based on the reflection and absorption of light. Pointing out that when you turn on the light in the room, the table receives all the colors, but only the blue color is reflected and reaches the observer's eyes, leading to the perception that the table is blue. Therefore, the learning objective focused on the relationship between light, color and visual perception.

Format 1, which is not aligned (see Figure 2a), violated the Principle of Coherence in the following sense: considering the learning objective mentioned above, by presenting the text in the material: "A peculiarity of colors is synesthesia... how to associate blue with a sweet taste", it deals with a curiosity, which does not favor the learning objective in question and may divert students' attention. For Mayer (2021), such types of text are understood as "seductive details", which can hinder the process of constructing the knowledge in question. Format 2 (see Figure 2b) proposes material that is in line with the Coherence Principle due to the absence of the aforementioned seductive detail, presenting only the content that is essential for understanding the subject.

**Figure 2a**. Format not in line with the Coherence Principle.

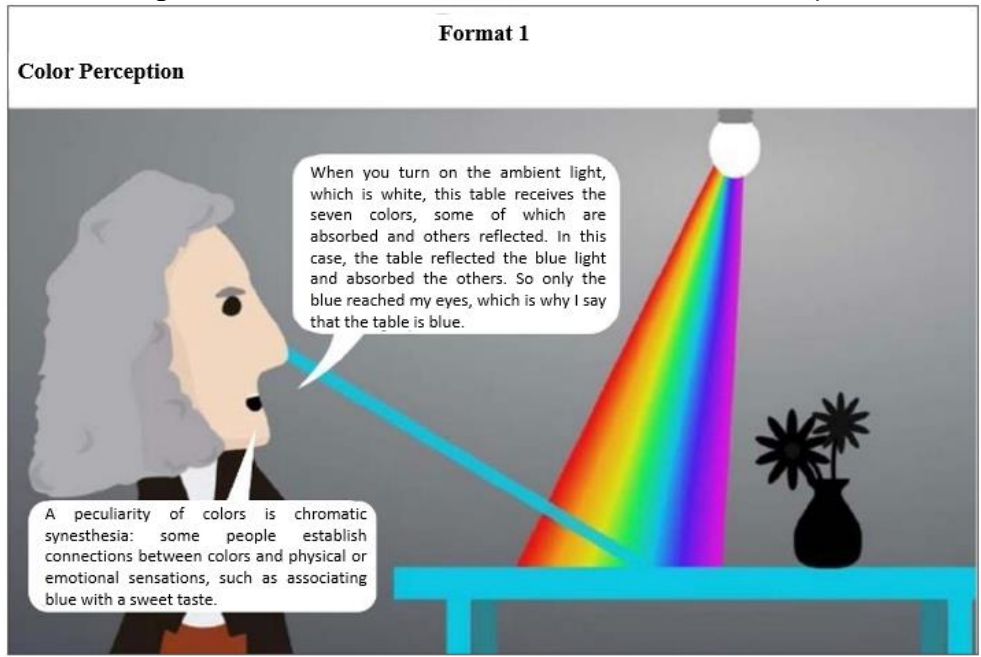

**Source:** adapted from Costa (2021, p. 23).

**Figure 2b.** Format aligned with the Coherence Principle.

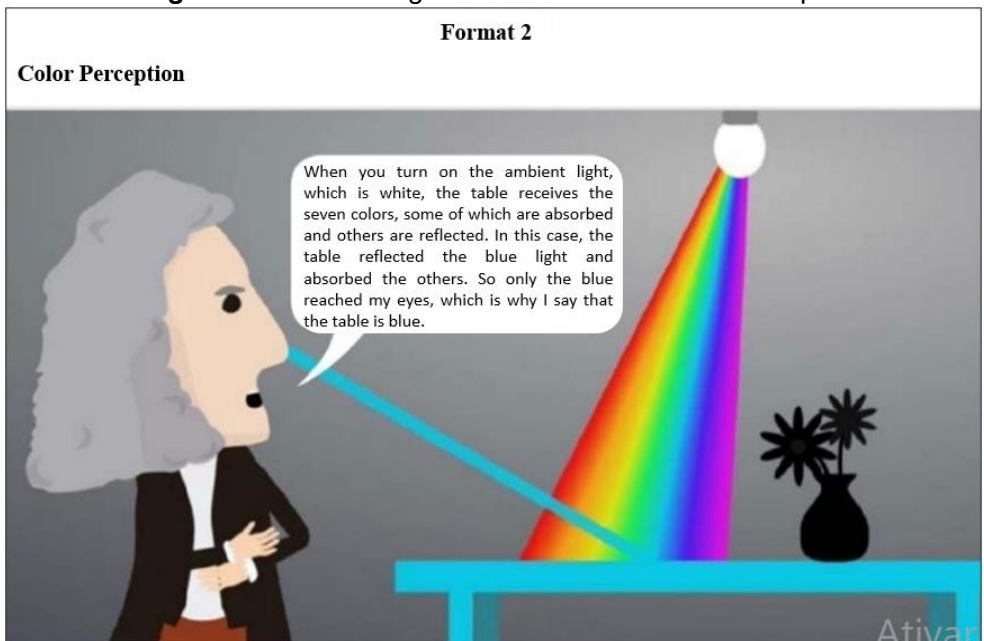

**Source:** adapted from Costa (2021, p. 23).

*Signaling Principle: selection of presentation formats on the process of light breaking down as it passes through a prism*

In order to analyze perception in relation to the Principle of Signaling, the multimedia presentation formats were designed with the aim of "*helping people to understand how the phenomenon of light decomposition occurs when it passes through a prism*". This occurs when, on passing through a prism, a beam of white light disperses into the colors of the spectrum: red, orange, yellow, green, blue, indigo and violet.

Format 1 (see Figure 3a) was proposed with the intention of violating the Signaling Principle. This format was characterized by the lack of Signaling to highlight key words such as: "*white light*", "*little hole in the wall*", "*prism*", "*red*", "*orange*", "*yellow*", "*green*", "*blue*", "*indigo*" and "*violet*". With the lack of signage on the material, this format did not guide the reader's eye to the essential information in the figure.

In contrast, Format 2 (see Figure 3b), in line with the Signaling Principle, highlights the material's key words, as well as highlighting important points in the figure. The insertion of these clues draws attention to the essential information in the multimedia material, directing the reader's gaze and making the instructional message more objective, facilitating comprehension, which in turn contributes to achieving the learning objective.

**Figure 3a.** Format not in line with the Signaling Principle.

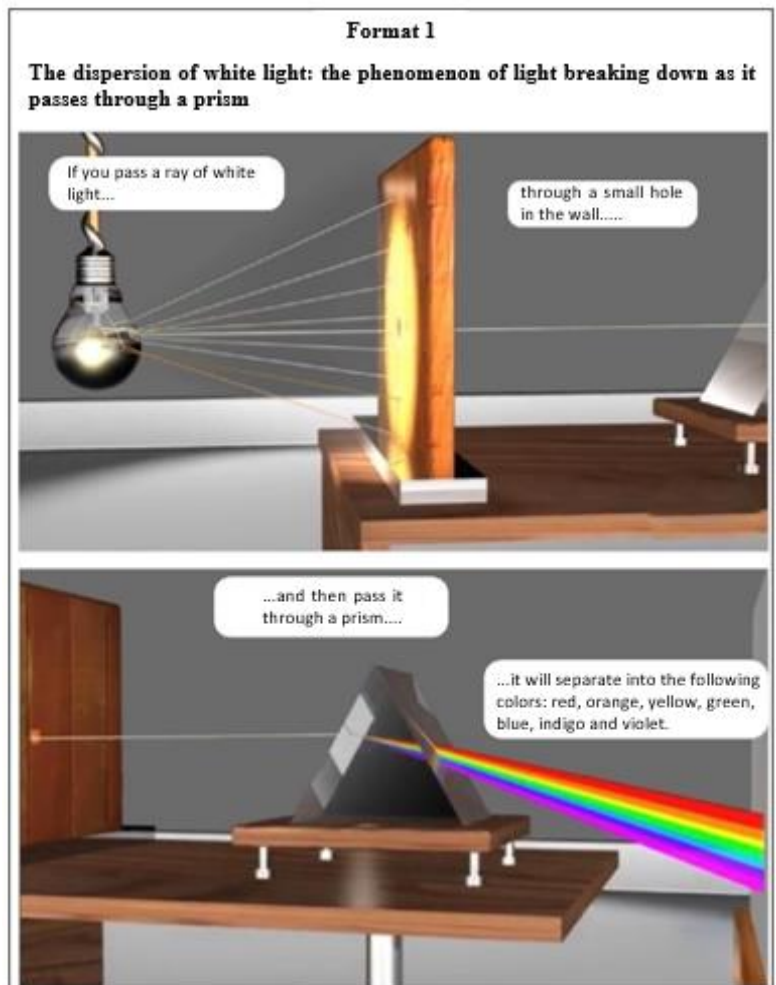

**Source:** adapted from Costa (2021, p. 2-3).

**Figure 3b.** Format in line with the Signaling Principle.

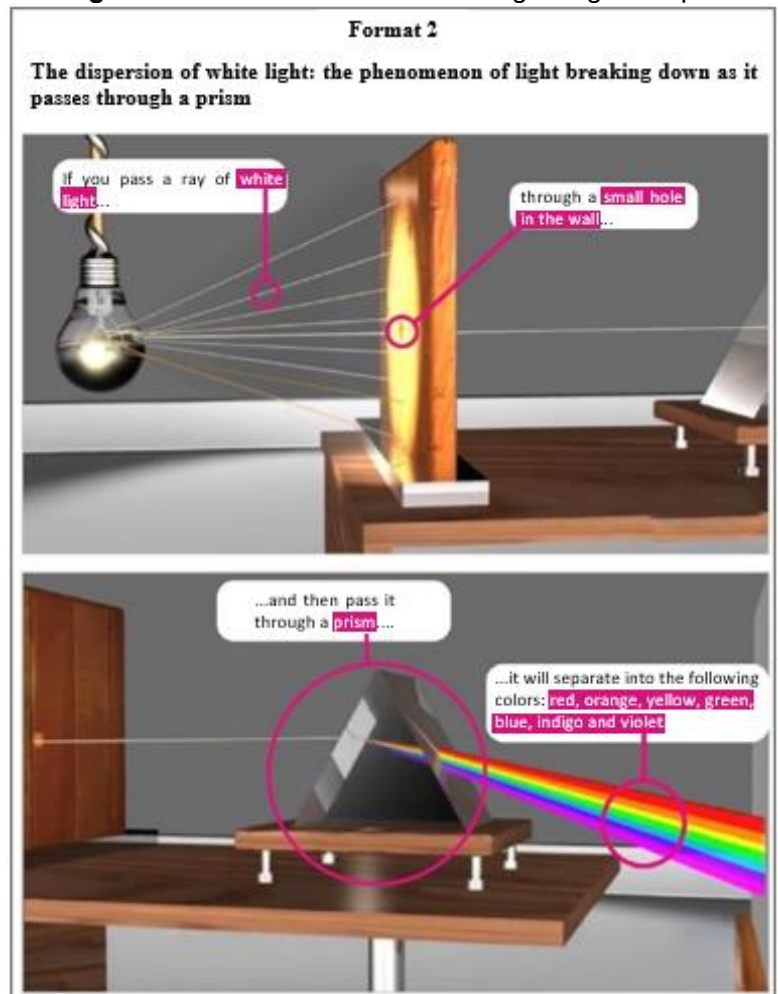

**Source:** adapted from Costa (2021, p. 2-3).

*Spatial Contiguity Principle: selection of presentation formats on primary and secondary light sources*

In order to analyze the perception of the Principle of Spatial Contiguity, the formats developed had the objective of "*helping to understand the concepts of primary and secondary sources of light*". Specifically, the learning objective in these presentations was to distinguish between primary sources - which emit their own light, such as the Sun and light bulbs - and secondary sources, which do not have their own light but shine when illuminated, such as the Moon and objects illuminated by external light sources. The formats dealt with the nature of these light sources and how they worked in relation to the emission and reflection of light.

Format 1 (see Figure 4a) is the format not aligned with the Spatial Contiguity Principle. It did not present the concepts of primary and secondary sources of light in a clear and objective way, as it displayed the texts on the content covered far away from the images. Format 2 (see Figure 4b), on the other hand, was proposed as a format in line with the principle in question. It highlighted the concepts clearly and objectively, presenting the images next to the corresponding text. This allowed text and image to be associated more quickly, making the content more accessible and understandable to students. Choosing teaching materials that take this principle into account allows students to understand the concepts presented more effectively, facilitating the learning process on the subject.

**Figure 4a.** Format not aligned with the Spatial Contiguity Principle.

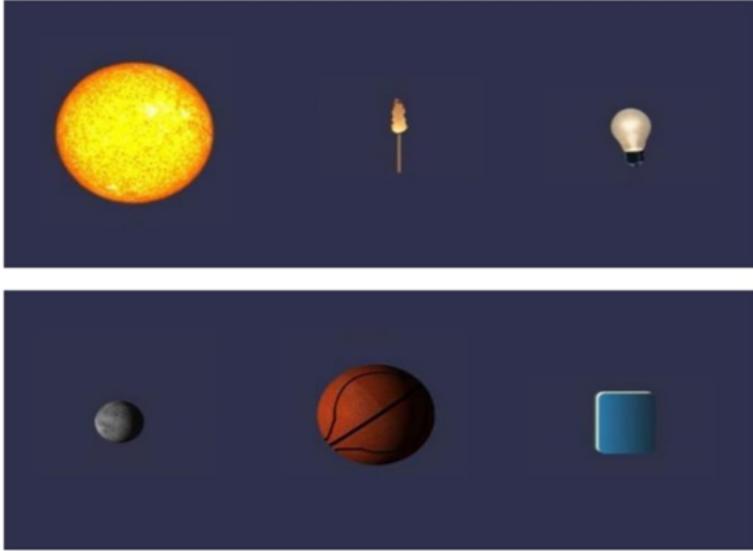

**Source:** adapted from Costa (2021, p. 17).

**Figure 4b.** Format aligned with the Spatial Contiguity Principle.

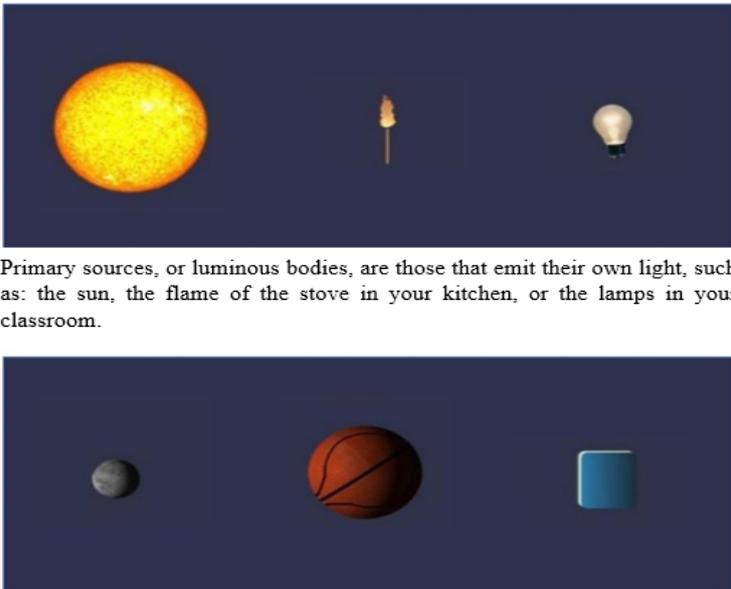

**Source:** adapted from Costa (2021, p. 17).

*Segmentation Principle: selection of presentation formats on the light reflection process*

With the aim of examining perception in relation to the Segmentation Principle, the formats had the objective of "*helping to understand the process of light reflection*". More specifically, the aim was to provide a detailed understanding of how light reflection occurs. This included explaining how light, when moving through a medium, interacts with a separating surface between two different media and, as a result, returns to propagate in the original medium. Therefore, the main focus was to convey the idea of how light behaves when it encounters an interface, which results in the phenomenon of light reflection.

Format 1 (see Figure 5a) was proposed in accordance with the Segmentation Principle, displaying the information in a segmented way, in parts, which enables the participant to understand the stages of the light reflection process. Each stage is accompanied by an image illustrating a specific moment, with the corresponding text just below the image. This organization into smaller segments can contribute to a clearer understanding of the whole, especially when it comes to students who are having their first access to the content (Mayer, 2021).

Format 2 (see Figure 5b) was proposed not in line with the principle mentioned, since it displays the phenomenon of light reflection in a single image and text, i.e. all the information is presented at once. The lack of segmentation, for students new to the subject, can generate cognitive overload and make learning difficult (Mayer, 2021).

**Figure 5a:** Format aligned with the Segmentation Principle

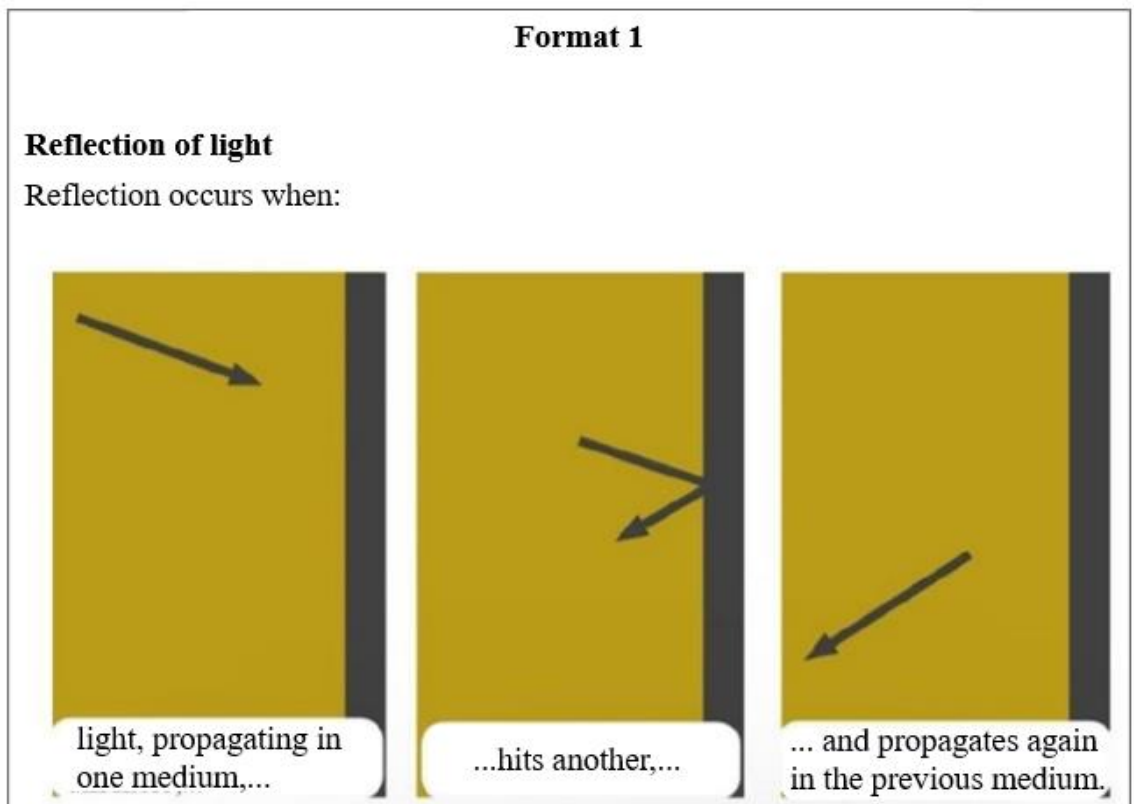

**Source:** adapted from Costa (2021, p. 10).

**Figure 5b.** Format not aligned with the Segmentation Principle.

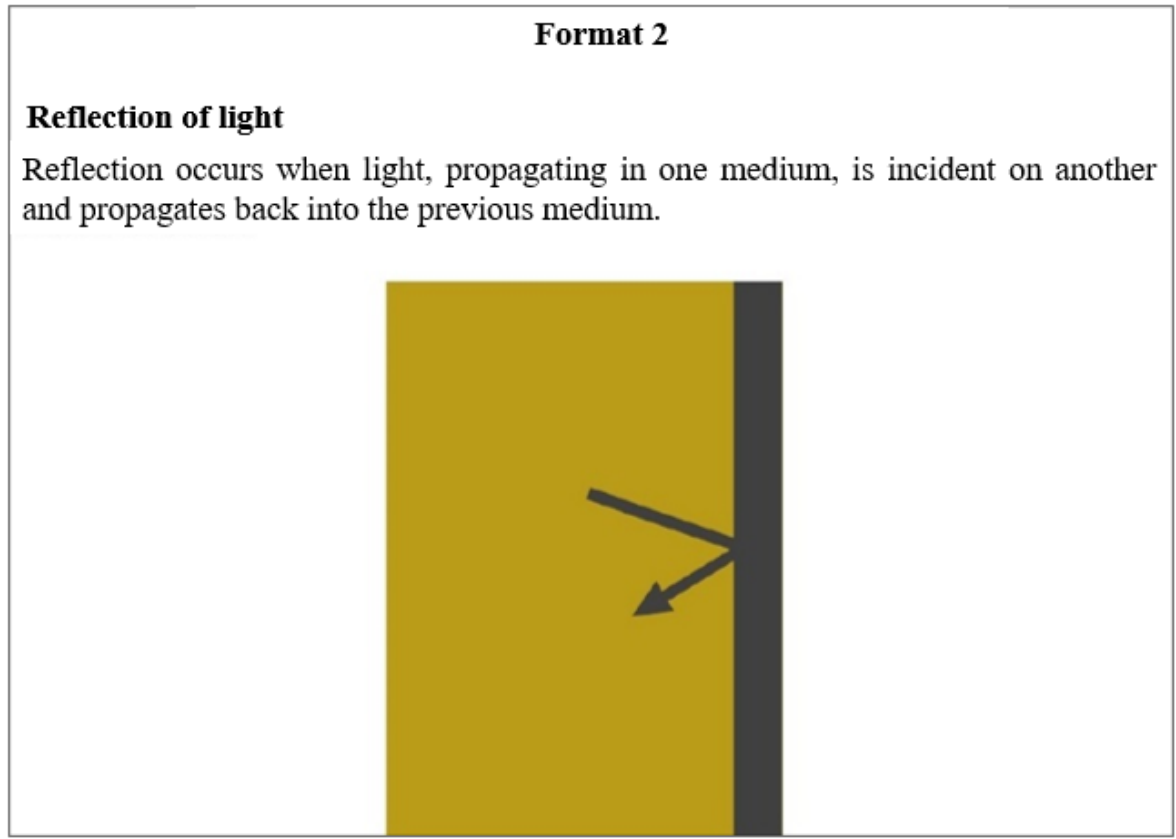

**Source:** adapted from Costa (2021, p. 10).

*Multimedia Principle: selection of presentation formats on the physical understanding of the umbra phenomenon*

To analyze the perception related to the Multimedia Principle, formats were developed with the aim of "*helping with the physical understanding of the umbra phenomenon*", as communicated to the participants. Specifically, the aim was to demonstrate how the umbra region is formed when direct sunlight is blocked by an object, such as a tree, resulting in a dark, unlit shadow.

Format 1 (see Figure 6a), in line with the Multimedia Principle, combines text and images to communicate information, eliminating the need for mental abstraction on the part of students. As already mentioned, in line with Mayer, the adoption of multimedia presentations brings significant advantages to the learning process. This allows students to simultaneously hold verbal and visual representations in their minds, opening up more opportunities to establish mental connections between these representations (Mayer, 2021). In contrast, Format 2 (see Figure 6b), which does not follow the Multimedia Principle, presents only text, leaving the reader with the task of imagining the phenomenon. Thus, Format 2 is unfavorable for the learning process, because when information is received exclusively in the form of text, performance may be lower, depending on the student's level of expertise in the subject.

**Figure 6a:** Format aligned with the Multimedia Principle.

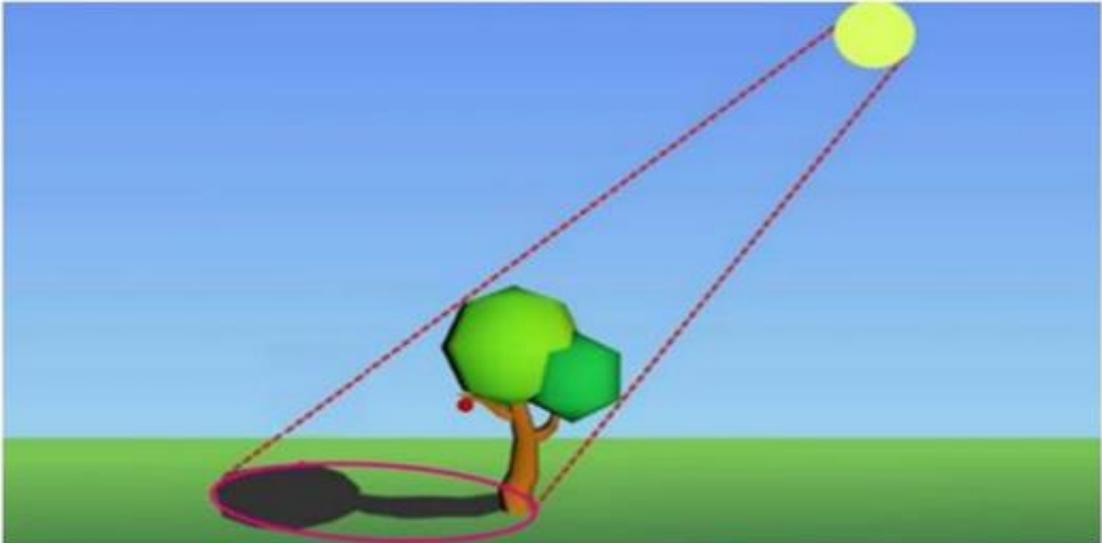

**Source:** adapted from Costa (2021, p. 50).

**Figure 6b.** Format not aligned with the Multimedia Principle.

**Source:** author's own production.

*Personalization Principle: Choosing Presentation Formats to Understand the Perception of Colors (Red, White and Black) through the Phenomena of Reflection and Absorption of Light*

In order to assess perception in relation to the Personalization Principle, formats were developed with the aim of "*helping students to understand the process of perceiving certain colors (red, white and black) through the phenomena of light reflection and absorption*". In particular, the learning objective of the formats was to explain the phenomenon of light reflection and absorption in colored objects. Using a simple example, the aim was to demonstrate how the color of an object is related to the way it interacts with light.

In order to assess perception in relation to the Personalization Principle, formats were developed with the aim of "helping students to understand the process of perceiving certain colors (red, white and black) through the phenomena of light reflection and absorption". In particular, the learning objective of the formats was to explain the phenomenon of light reflection and absorption in colored objects. Using a simple example, the aim was to demonstrate how the color of an object is related to the way it interacts with light.

Format 1 (see Figure 7a), in line with the Personalization Principle, adopts a dialogical approach to presenting information, establishing a direct dialogue between the author and the reader through less formal language. This format actively seeks to promote the engaged participation of the reader, since the incorporation of a conversational style aims to stimulate student involvement and facilitate their understanding of the teacher's messages. According to Mayer (2021), this interaction leads to an increase in cognitive processing activity on the part of the student, since they are more dedicated to selecting, organizing and integrating the information presented to them. Format 2 (see Figure 7b) is not aligned with the Personalization Principle, as it does not actively promote reader participation. This format consists of a text that adopts formal language and does not involve dialogue, which, in the context of the Personalization Principle, does not stimulate student engagement and can make understanding the content more difficult.

**Figure 7a.** Format aligned with the Personalization Principle.

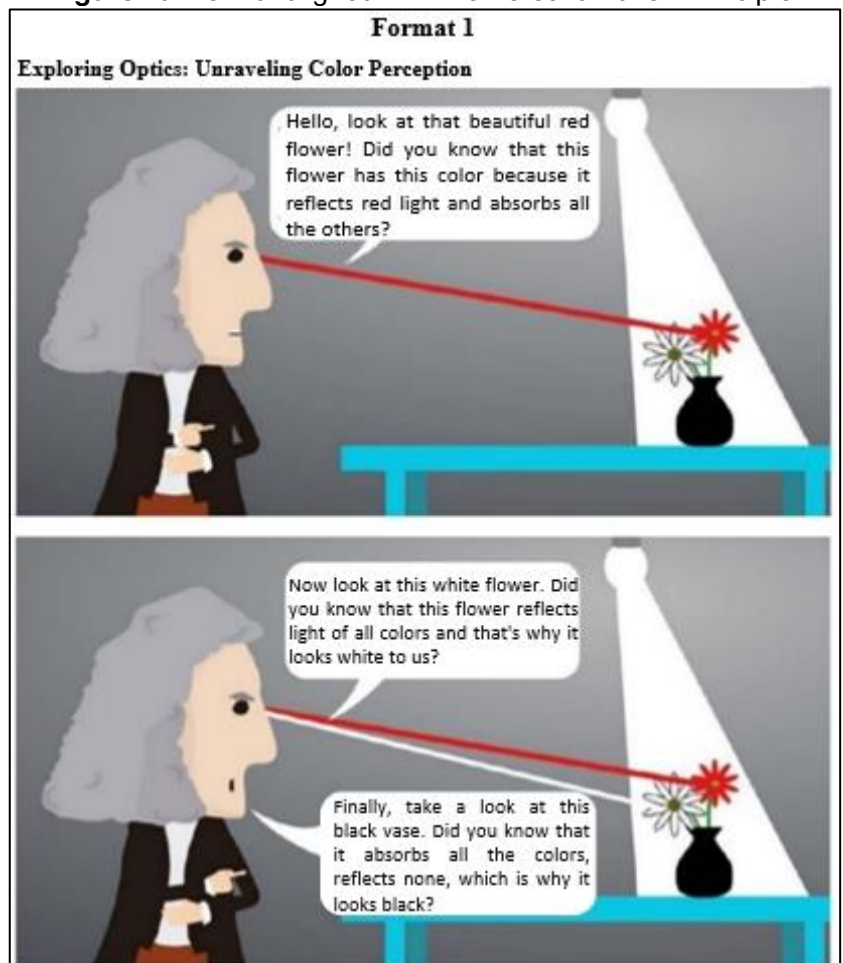

**Source:** adapted from Costa (2021, p. 26).

**Figure 7b.** Format not in line with the Personalization Principle.

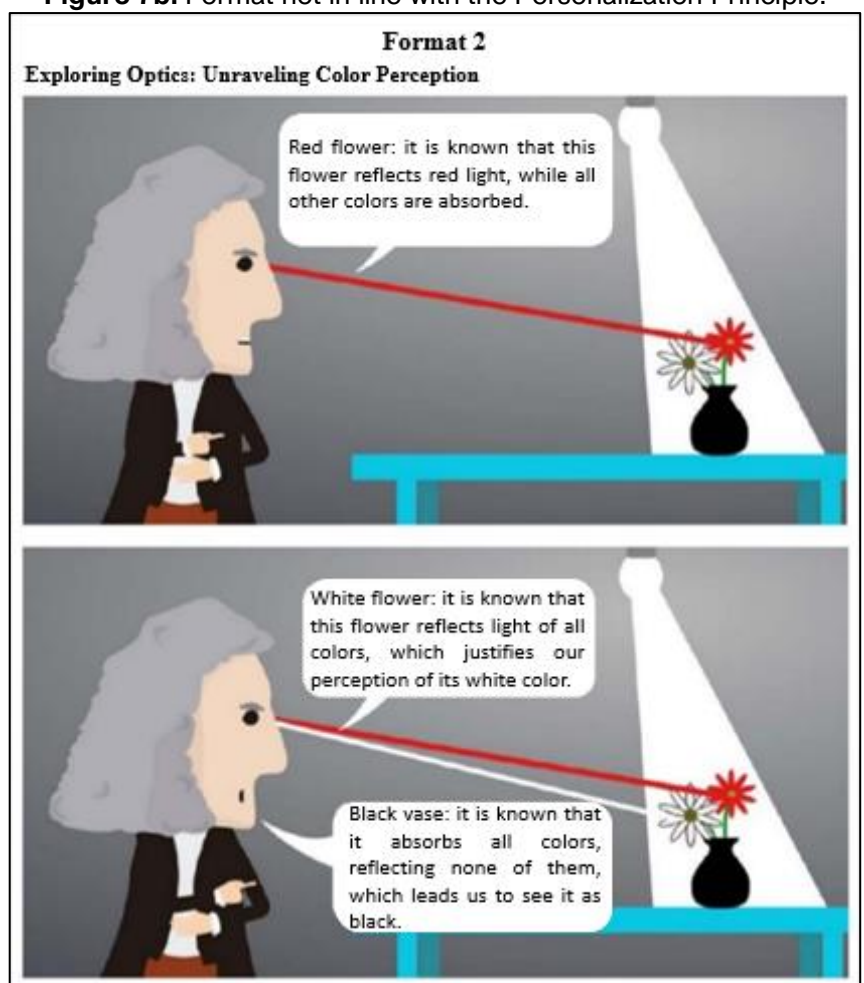

**Source:** adapted from Costa (2021, p. 26).

## 3.3 Specific Aspects of Stage 2/2: Validation Questionnaires

In order to prevent the results of this research from being prejudiced by purely random selections, or those whose motivations were not related to CTML, validation questionnaires were introduced to rule out any random/unrelated selections, validating only those selections whose motivations were related to CTML.

The validation questionnaires began with the following text: "*On the following two sheets, you will find two teaching materials, one on each sheet, which explain the same topic, but in different formats. Imagine that you need to select one of these materials to teach a lesson, and you have to choose which format to use. Take into account which one will provide more efficient learning, in the sense of helping the student to understand the process...*". The process mentioned was specific to each material. All the validation questionnaires also had questions in common, one objective and two subjective:

(a) The first question was objective in nature and consisted of the following question: "*Which teaching material do you think would result in better learning for your student?*", and had two answer options: "*Teaching material with Format 1*" or "*Teaching material with Format 2*". The aim of this question was to determine whether or not the participants would choose material aligned with a specific CTML principle.

(b) The second question was subjective in nature and involved the following question: "*Why do you believe that the teaching material you have chosen has a better format than the other? Answer the question, highlighting the characteristics in the material that led you to make that choice.*" The aim of this question was to see whether the participants' choice was accompanied by justifications that were close to the principles of CTML. Participants were asked to explain why they believed that the material they chose had a superior format compared to the other. The answers provided information on the specific characteristics of the materials that influenced the participants' choice, and it was possible to identify the extent to which the principles of CTML were covered or not in these answers.

(c) The third question was also subjective in nature and asked: "*Indicate which characteristics of the teaching material you didn't choose make it inferior to the material you did choose*". By asking the participants to highlight the specific characteristics of the material they didn't choose that made it inferior, it was possible to identify the characteristics that negatively influenced their decisions. The aim of the question, therefore, was to confirm the extent to which the principles of CTML were taken into account or not in the participants' choices.

The choice of a material format aligned with a given CTML principle [item (a), above] was considered valid when the following conditions were met: the participant justified their choice in item (b) with arguments aligned with the corresponding CTML principle; in addition, the participant indicated in item (c) the characteristics aligned with the corresponding CTML principle that were absent in the teaching material that was not chosen.

## 3.4 Specific aspects of Stage 3: Final Questionnaire

The main aim of the final questionnaire was to identify the profile of the participants and check whether they had any prior knowledge of the principles of CTML. The central question of the questionnaire was: "*Do you know or have you heard of the Cognitive Theory of Multimedia Learning and its Principles? More specifically, the Coherence Principle, the Signaling Principle, the Spatial Contiguity Principle, the Segmentation Principle, the Multimedia Principle or the Personalization Principle?*" This question aimed to identify whether participants had any knowledge of CTML and its principles, ensuring that only data from those without prior knowledge was included in the analysis. Participants who indicated familiarity with CTML had their data excluded to avoid compromising the validity of the results, since their answers might not have been spontaneous. In addition, the questionnaire collected information on the profile of the participants, such as name, age, period of undergraduate study and teaching experience. For the teachers, an additional piece of information was requested: whether they were studying for a postgraduate degree.

## 4. RESULTS AND DISCUSSION

## 4.1 Results of Stage 2/1

In this section, we present the results that refer exclusively to the participants' choices of multimedia presentation formats, with the aim of identifying how many of them opted for formats aligned with the CTML principles. With regard to the Principle of Coherence, it can be seen that only 12.5% of students opted for material aligned with this principle, while among teachers this option was 60% (see Table 1). A possible explanation for this discrepancy is that the students would see the "seductive detail" (interesting visual and/or verbal element that is more irrelevant to learning) as an element that, instead of getting in the way, would motivate the student to learn; whereas the teachers, because of their teaching experience, would see the seductive detail as an irrelevant element that would get in the way of the learning process.

With regard to the Principles of Signaling and Spatial Contiguity, 62.5% and 75% of students opted for materials in line with these principles, while among teachers this option was 80% (see Table 1). A possible explanation for this result also lies in the accumulated teaching experience of the teachers, who would better understand the importance of Signaling and the need to bring texts close to the images.

With regard to the Segmentation Principle, 100% of students opted for materials aligned with this principle, while 60% of teachers did (see Table 1). A possible explanation for this result lies in the different expertise of students and teachers in relation to the subject: students, because they have less expertise, would identify segmented material as better for learning, while teachers, influenced by their expertise, would no longer perceive the need for segmentation.

With regard to the Multimedia Principle, 100% of students and teachers opted for material in line with this principle (see Table 1). This indicates that it is unanimously clear that it is necessary to associate figures with the text in order to improve learning.

With regard to the Personalization Principle, 75% of students opted for materials in line with this principle, while 80% of teachers did (see Table 1). This indicates that the majority of these participants see the need for more dialogic materials (less formal language) to improve learning.

**Table 1.** Number of participants who chose the format aligned with CTML principles (%)

| CTML principles | Physics teachers (5 Participants) | Physics undergraduates (8 Participants) |
|---|---|---|
| Coherence | 3/5 (60%) | 1/8 (12.5%) |
| Signaling | 4/5 (80%) | 5/8 (62.5%) |
| Spatial contiguity | 4/5 (80%) | 6/8 (75%) |
| Segmentation | 3/5 (60%) | 8/8 (100%) |
| Multimedia | 5/5 (100%) | 8/8 (100%) |
| Personalization | 4/5 (80%) | 5/8 (62.5%) |

## 4.2 Results of Stage 2/2

In this section, we exemplify the execution of Stage 2/2 of our study, which aimed to validate the participants' choices regarding multimedia presentation formats.

*Example 1: Validated choice*

Criteria: This occurs when the participant chooses a material format that is aligned with a CTML principle and their justification is also aligned with the essential elements of that principle's definition.

Description of the situation: The following choice of a material format aligned with the Signaling Principle was considered valid. The participant justified his choice [item (b) of Section 3.3] with the following argument: "*Eye-catching colors to highlight words and physical phenomena help with didactics.*" This argument is clearly aligned with the aforementioned CTML principle. In addition, the participant indicated [item (c) of Section 3.3] the characteristics aligned with the corresponding CTML principle that were absent in the teaching material that was not chosen: "*Only explained, and the effects are not highlighted.*"

*Example 2: unvalidated choice - situation 1*

Criteria: Occurs when the participant chooses a material format in accordance with a CTML principle, but their justification does not come close to the essential elements of the principle's definition.

Description of the situation: The following choice of a material format aligned with the Coherence Principle was not considered valid. The participant justified his choice [item (b) of Section 3.3] with the following argument: "*Because it is more direct and easier to understand the explanation compared to the other. The choice was due to the clarity of the explanation.*" This argument is in line with the aforementioned CTML principle. However, the participant did not indicate [item (c) of Section 3.3] the characteristics aligned with the corresponding CTML principle that were absent in the didactic material that was not chosen: "*Light bulb, dark vessel that is not illuminated by light, the same one that is totally blue and the colors emitted by the light when it rises.*" In other words, the participant did not identify the seductive detail, included in the material to violate the Principle of Coherence, as bad for learning.

*Example 3: unvalidated choice - situation 2*

Criteria: When the participant did not choose a material format aligned with a CTML principle, and their justification was not aligned with the essential elements of the principle's definition.

Description of the situation: The following choice of a material format not in line with the Signaling Principle was not considered valid. The participant justified his choice [item (b) of Section 3.3] with the following argument: "*I believe that material 1 is in a less conventional format and in harmony with the colors, which makes information stand out.*" This argument is not in line with the CTML principle. In addition, the participant did not indicate [item (c) of Section 3.3], in the didactic material that was not chosen, the characteristics aligned with the corresponding CTML principle that were absent in that material: "*The spelled words and circles with very intense colors hinder the focus on the information.*"

*Example 4: unvalidated choice - situation 3*

Criteria: When the participant has not chosen a material format in line with a CTML principle, and has not given a coherent justification for this choice.

Description of the situation: The following choice of a material format not aligned with the Personalization Principle was not considered valid. The participant justified his choice [item (b) of Section 3.3] with the following argument: "*Material 2, because it addresses the whole context of optics, affirming knowledge*.", an argument that is not aligned with the aforementioned CTML principle. The participant also did not indicate [item (c) of Section 3.3], in the didactic material that was not chosen, the characteristics aligned with the corresponding CTML principle that were absent in that material: "*Material 1 manages to meet both needs, as well as creating a 'questioning' for the student. I would use both materials*".

Thus, after analyzing the justifications given by the participants, it was possible to validate the data in Table 1 and produce a new table of results, Table 2, which will be discussed below.

**Table 2.** Number of participants who chose and justified the format in line with CTML principles (%)

| CTML principles | Physics teachers (Originally 5 Participants) | Physics undergraduates (Originally 8 Participants) |
| --- | --- | --- |
| Coherence | 3/5 (60%) | 0/7 (0%) |
| Signaling | 4/5 (80%) | 5/8 (62.5%) |
| Spatial contiguity | 4/5 (80%) | 6/7 (85.71%) |
| Segmentation | 3/5 (60%) | 8/8 (100%) |
| Multimedia | 5/5 (100%) | 8/8 (100%) |
| Personalization | 4/4 (100%) | 5/6 (71,4%) |

## 4.3 Results of Stage 3

In this section, we present the results of Stage 3, which involved analyzing the profile of the participants and their knowledge of CTML. The sample consisted of 13 participants, including physics teachers and undergraduates, who said they did not know the principles of CTML.

The majority of the students were between 20 and 22 years old, six were in their fifth term and two were in their seventh term of the Physics degree course. Half of these students reported experience in teaching, especially in tutoring and as scholarship holders in pedagogical residency projects, but only three had specific experience in teaching physics.

The teachers ranged in age from 22 to 51. Of the five participants, three were studying for a master's degree, while two already had a master's degree, and one of them also had a post-doctorate. Four of the teachers had experience in teaching physics, working in both primary and higher education. Only one of the teachers said he had no experience of teaching physics.

## 4.4 Results of Stage 4

In this section, we present three aspects of Table 2. The first aspect of Table 2 is the fact that, despite the fact that the participants reported that they had no formal instruction in CTML, all the participants chose materials aligned with the Multimedia Principle and, what's more, the majority chose formats aligned with the principles of Signaling, Spatial Contiguity, Segmentation and Personalization. There was, however, a not insignificant proportion who did not select materials aligned with these principles, the most extreme case being that no licensee chose material aligned with the Coherence Principle.

The second aspect is that teachers chose and justified materials aligned with the principles of Coherence, Signaling and Personalization more than undergraduates. For example, in the case of the Coherence Principle, 60% of teachers chose and adequately justified the format of material aligned with this principle, while none of the undergraduates made the same choice (as already mentioned).

To understand these two aspects observed in the results of Table 2, we used Ausubel's (2000) Significant Learning Theory, more specifically, the concepts of subsumers and advanced organizers. According to Ausubel (2000), subsumers or prior knowledge are pre-existing knowledge structures developed from previous experiences. Ausubel proposed the term advanced organizer to describe materials that could be used to activate this prior knowledge and facilitate meaningful learning. These advanced organizers are presented before learning new content and have the function of connecting pre-existing knowledge with new information, providing a deeper and more effective understanding (Driscoll, 2014). Based on these concepts, we developed explanatory hypotheses for the two aspects observed in Table 2.

With regard to the first aspect, which deals with the choice of teaching materials aligned with the principles of CTML, it is possible that, even without formal knowledge of a given principle, the participant has become familiar with materials aligned with that principle throughout their professional practice. In this way, they may have previously accumulated pedagogical knowledge (prior knowledge) related to a given principle, which may have been activated during the research, allowing the participant to identify and select the most appropriate material format. This possible activation of prior knowledge would take place through advanced organizers provided by the comparison between formats aligned and not aligned with a given principle.

As for the second aspect, it can be explained by the absence or non-activation of prior knowledge (pedagogical knowledge) accumulated as a result of didactic experience related to the principle. Even if the participants had such accumulated knowledge, the advanced organizer provided by the comparison between formats may not have been enough to activate it. The questionnaires, which asked for justifications for the choices, helped to identify whether prior knowledge had been activated. In the case of the Principle of Coherence, the lack of activation of this knowledge may have contributed to undergraduates not recognizing the importance of excluding irrelevant information and not identifying the material most in line with the principle.

The third aspect is that undergraduates chose and justified materials aligned with the Segmentation Principle more than teachers. One possible explanation may lie in the different expertise between students and teachers on the subject: graduates, with less formal knowledge, tend to consider segmented materials more suitable for learning, while teachers, influenced by their greater expertise, may not perceive the same need for segmentation. According to Bransford, Brown and Cocking (2000, p. 44), "expertise can sometimes hinder teaching, because many experts forget what is easy and what is difficult for students".

It is also worth mentioning an interesting convergence between the results obtained by Colombo and Antonietti (2006), regarding the principles of Spatial Contiguity and Coherence, and those obtained in this study. In relation to these principles, these authors obtained the following results: 88.4% and 69.0%, respectively, were the percentages of choices in agreement with these two principles. Our results indicated 80% (teachers) and 85.71% (undergraduates) respectively for the Spatial Contiguity Principle, and 60% (teachers) and 0% (undergraduates) for the Coherence Principle. This interesting convergence (with the exception of the graduate students' result for the Principle of Coherence), however, should be considered with reservations, since Colombo and Antonietti (2006) carried out a statistical survey, while the present work consisted of a case study. Further research will be needed to confirm these convergences.

## 5. FINAL CONSIDERATIONS

As mentioned, the results corresponding to the principles of Spatial Contiguity and Coherence, obtained in this case study involving 13 participants from the field of Physics, proved to be convergent with those obtained by Colombo and Antonietti (2006), involving 112 participants from the fields of Administration, Psychology, Literature and Arts. This convergence in these two principles (the only principles in common in the two studies) supports a possible generalization of the results obtained here for analogous situations (case studies with other participants from the Physics area, as well as from other areas), authorizing the expectation that future studies can generalize the results to a larger and more representative number of participants.

In addition, our research revealed that the presentation format aligned with the Multimedia Principle was chosen by all participants, while the formats aligned with the Principles of Signaling, Spatial Contiguity, Segmentation and Personalization were chosen by the majority. Thus, even without formal instruction on CTML, physics teachers and students chose materials aligned with these principles. However, a significant proportion of participants did not choose materials aligned with these last four principles. In addition, material aligned with the Principle of Coherence was not chosen by any physics undergraduates. Therefore, although CTML has been widely studied over the last four decades and its principles are fundamental to the development of effective instructional materials, our research reveals the need to include formal instruction on CTML in university physics degree courses. This would ensure greater awareness and application of these principles to improve the quality of teaching and learning in the choice, use and development of multimedia teaching materials.

## ACKNOWLEDGMENTS

This work was supported by the Coordination for the Improvement of Higher Education Personnel (CAPES) - Brazil, Funding Code 001, and also by the National Council for Scientific and Technological Development (CNPq) - Brazil, Process 408735/2023-6 CNPq/MCTI.